\documentclass[12pt]{article}
\usepackage{graphicx}

\usepackage{xspace}
\usepackage{amsmath}
\usepackage{calc}

\def\sPlot{\mbox{\em sPlot}}

\def\lhcb {\mbox{LHCb}\xspace}

\def\CP {{CP}\xspace}

\def\babar  {\mbox{BaBar}\xspace}
\def\belle  {\mbox{Belle}\xspace}

\def\PB      {\ensuremath{B}\xspace}                 
\def\PD      {\ensuremath{D}\xspace}                 
\def\PK      {\ensuremath{K}\xspace}                 
\def\Ps      {\ensuremath{s}\xspace}         
\def\Pb      {\ensuremath{b}\xspace}         
\def\Pc      {\ensuremath{c}\xspace}         
\def\Ppi         {\ensuremath{\pi}\xspace}                 
        
\def\squark    {\ensuremath{\Ps}\xspace}
\def\bquark    {\ensuremath{\Pb}\xspace}
\def\cquark    {\ensuremath{\Pc}\xspace}

\def\pion  {\ensuremath{\Ppi}\xspace}
\def\pip   {\ensuremath{\pion^+}\xspace}

\def\pim   {\ensuremath{\pion^-}\xspace}

\def\kaon  {\ensuremath{\PK}\xspace}
\def\K       {\ensuremath{\PK}\xspace}
\def\Kp    {\ensuremath{\kaon^+}\xspace}
\def\Km    {\ensuremath{\kaon^-}\xspace}
\def\Kpm     {\ensuremath{\K^\pm}\xspace}

\def\Dbar    {\kern 0.2em\overline{\kern -0.2em \PD}{}\xspace}

\def\B       {\ensuremath{\PB}\xspace}
\def\Bbar    {\kern 0.18em\overline{\kern -0.18em \PB}{}\xspace}

\def\Bzb     {\ensuremath{\Bbar^0}\xspace}

\def\Bpm     {\ensuremath{\B^\pm}\xspace}

\def\Bsb     {\ensuremath{\Bbar^0_\squark}\xspace}
\def\Bdb     {\ensuremath{\Bbar^0}\xspace}
  
\def\Bnb     {\ensuremath{\Bbar^0_{(\squark)}}\xspace}  

\def\jpsi  {\ensuremath{J/\psi}\xspace}

\def\BsJphi     {\ensuremath{\Bsb \to J/\psi \phi}\xspace}

\def\BsJKK     {\ensuremath{\Bsb \to J/\psi \Kp \Km}\xspace}
\def\BsJpipi     {\ensuremath{\Bsb \to J/\psi \pip \pim}\xspace}

\def\BzJKK     {\ensuremath{\Bzb \to J/\psi \Kp \Km}\xspace}
\def\BzJpipi     {\ensuremath{\Bzb \to J/\psi \pip \pim}\xspace}
\def\BXK         {\ensuremath{\Bpm \to X(3872) \Kpm}\xspace}
\def\BnJKK     {\ensuremath{\Bnb \to J/\psi \Kp \Km}\xspace}
\def\BnJpipi     {\ensuremath{\Bnb \to J/\psi \pip \pim}\xspace}
\def\BnJhh    {\ensuremath{\Bnb \to J/\psi h^+ h^-}\xspace}
\def\BppK         {\ensuremath{\Bpm \to p\overline{p} \Kpm}\xspace}

\def\pp      {\ensuremath{pp}\xspace}

\def\xff {\ensuremath{X(3872)}\xspace}
\def\pipi{\ensuremath{\pi\pi}\xspace}
\def\hh{\ensuremath{h^+h^-}\xspace}
\def\xppj{\ensuremath{\xff \to J/\psi \pi^+\pi^- }\xspace}
\def\jmm{\ensuremath{\jpsi\to \mu^+\mu^-}\xspace}

\def\jpc{\ensuremath{J^{PC}}\xspace}

\newcommand{\mpp}{M_{p \bar p} < 2.85}

\def\pbnr{}
\def\speaker{Paras Naik}
\def\onbehalfof{the \lhcb Collaboration}
\def\title{Studies of charmed states in amplitude analyses at \lhcb}
\def\affiliation{School of Physics\\
University of Bristol, Bristol, United Kingdom}
\def\support{The workshop was supported by the University of Manchester, IPPP, STFC, and IOP}

\newcommand\pubnumber{\pbnr}
\newcommand\pubdate{\today}
\def\Title#1{\begin{center} {\Large #1 } \end{center}}
\def\Author#1{\begin{center}{ \sc #1} \end{center}}

\newcommand{\OnBehalf}[1]{\sbox0{#1}\ifdim\wd0=0pt
        {}
	\else
	{\\on behalf of #1}
	\fi}
\newcommand{\SupportedBy}[1]{\sbox0{#1}\ifdim\wd0=0pt
        {}
	\else
	{\footnote{#1}}
	\fi}
\def\Address#1{\begin{center}{ \it #1} \end{center}}

\newcommand\pubblock{\includegraphics[width=5cm]{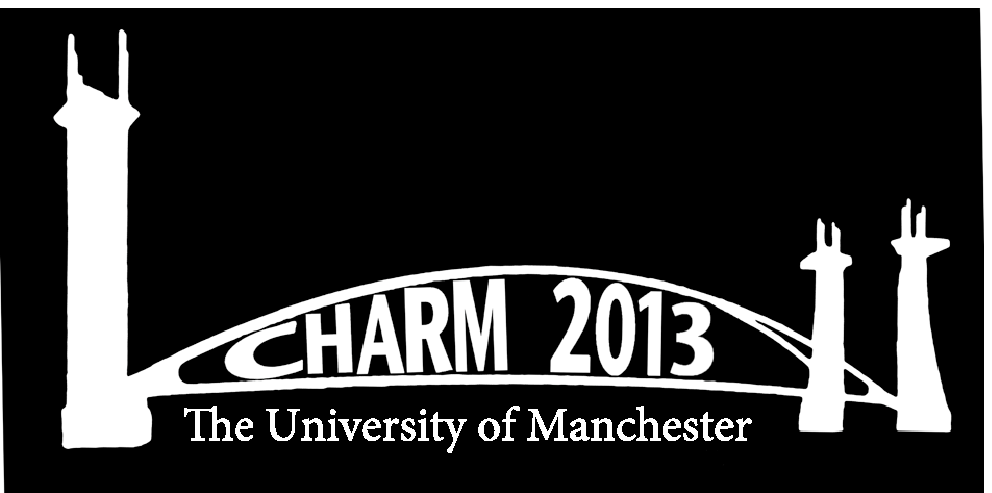}\hfill{\begin{tabular}{l} \pubnumber\\
         \pubdate  \end{tabular}}}
\newenvironment{Abstract}{\begin{quotation}  }{\end{quotation}}

\newenvironment{Presented}{\begin{quotation}\begin{center}~\\PRESENTED AT\end{center} \bigskip \begin{center} \begin{large} }       { \end{large} \end{center} \end{quotation} }
      
\def\Acknowledgements{\bigskip  \bigskip \begin{center} \begin{large}
\bf ACKNOWLEDGEMENTS \end{large}\end{center}}
\def\venue{The 6$^{th}$ International Workshop on Charm Physics\SupportedBy{\support}\\
(CHARM 2013)\\
Manchester, UK,  31 August -- 4 September, 2013}

\def\beq{\begin{equation}}
\def\eeq#1{\label{#1}\end{equation}}
\def\eeqn{\end{equation}}

\def\beqa{\begin{eqnarray}}
\def\eeqa#1{\label{#1}\end{eqnarray}}
\def\eeqan{\end{eqnarray}}

\let\bar=\overbar

\def\L{{\cal L}}
\def\M{{\cal M}}

\def\Dslash{\not{\hbox{\kern-4pt $D$}}}
\def\dslash{\not{\hbox{\kern-2pt $\del$}}}

\def\msb{{\bar{\ssstyle M \kern -1pt S}}}

\begin{document}
\begin{titlepage}
\pubblock

\vfill
\Title{\title}
\vfill
\Author{\speaker\OnBehalf{\onbehalfof}}
\Address{\affiliation}
\vfill
\begin{Abstract}
Amplitude analysis is a powerful tool to study the properties of intermediate resonances produced in the decays of $B$ mesons. 
At \lhcb we have studied \BXK, where $\xppj$, to determine the quantum numbers of the \xff, and \BppK to learn more about $(c\overline{c}) \to p\overline{p}$ transitions. We also exploit the spin of the $J/\psi$ to perform amplitude analyses of the decays \BnJpipi  and \BnJKK.
Our results use 1.0 fb$^{-1}$ of data taken in 2011 from $\sqrt{s}=7$ TeV \pp collisions, provided by the LHC.
\end{Abstract}
\vfill
\begin{Presented}
\venue
\end{Presented}
\vfill
\end{titlepage}
\def\thefootnote{\fnsymbol{footnote}}
\setcounter{footnote}{0}


\section{Introduction}

The dominant weak decay $\bquark \to \cquark$ results in the production of charmed states in $B$ meson decays that can be well-explored by the \lhcb experiment. The following is a summary of how exploiting charmed states in $B$ decays at \lhcb has lead to a better understanding of both. Our results use 1.0 fb$^{-1}$ of data from $\sqrt{s}=7$~TeV \pp collisions, provided by the LHC in 2011. This data was collected by the \lhcb detector (described in Refs. \cite{Alves:2008zz,Aaij:2012me}).

\section{Measurement of the \xff quantum numbers via \BXK decays \cite{BXK}}

The 
\xff  
was discovered in $B^+$ decays\footnote{The inclusion of charge-conjugate states is implied in this proceeding.}
by the \belle experiment in 2003
\cite{Choi:2003ue}, and has been confirmed by several
other experiments \cite{CDFPhysRevLett.93.072001,D0Abazov:2004kp,BaBarPhysRevD.71.071103}.
\xff production has most recently been studied at the LHC \cite{Aaij:2011sn,Chatrchyan:2013cld},
however the nature of the \xff remains unclear. The open explanations for this state are 
conventional charmonium and exotic states such as 
$D^{*0}\bar{D}{}^0$ molecules \cite{Tornqvist:2004qy}, 
tetra-quarks \cite{Maiani:2004vq}, or their mixtures \cite{Hanhart:2011jz}.
To determine the best explanation, we need to determine the quantum numbers  $J$ (total angular momentum), $P$ (parity), 
and $C$ (charge-conjugation) of the \xff.

The CDF experiment analyzed three-dimensional (3D) angular correlations in a sample of inclusively-reconstructed $\xppj$ , $\jmm$ decays  \cite{Abulencia:2006ma}. 
A $\chi^2$ fit of $\jpc$ hypotheses to the binned 3D distribution of the $\jpsi$ and $\pipi$ helicity angles ($\theta_{\jpsi}$, $\theta_{\pipi}$)
\cite{Jacob:1959at,Richman:1984gh,PhysRevD.57.431},
and the angle between their decay planes, excluded all spin-parity assignments except for $1^{++}$ or $2^{-+}$.
The \belle collaboration 
concluded that their data were equally well described by the $1^{++}$ and $2^{-+}$ hypotheses, by studying one-dimensional distributions in three different angles \cite{Choi:2011fc}.
The \babar experiment used information from
$\xff\to\omega\jpsi$, 
$\omega\to\pi^+\pi^-\pi^0$ events to favor the $2^{-+}$ hypothesis, which had a confidence level ({\rm CL}) of $68\%$, 
over the $1^{++}$ hypothesis, but the latter was not ruled out ({\rm CL}~$=7\%$)  \cite{delAmoSanchez:2010jr}.

The angular correlations in the $B^+$ decay  
carry significant information about the $\xff$ quantum numbers.
A first analysis of the complete five-dimensional angular correlations
of the $B^+\to\xff K^+$, $\xppj$, $\jmm$ decay chain has been performed
using the 2011 \lhcb data sample. A fit to the data yields $313\pm26$ $\BXK$ candidates.   
To discriminate between the $1^{++}$ and $2^{-+}$ assignments we
use the likelihood-ratio test, 
which in general provides the most powerful test 
between two hypotheses \cite{james2006statistical}. 
The probability distribution function (PDF) for each $\jpc$ hypothesis, $J_X$,  is defined
in the 5D angular space $\Omega\equiv$
$(\cos\theta_X,\cos\theta_{\pipi},\Delta\phi_{X,\pipi},\cos\theta_{\jpsi},\Delta\phi_{X,\jpsi})$
by the normalized product of
the expected decay matrix element ($\M$) squared
and of the reconstruction efficiency ($\epsilon$), 
${\rm PDF}(\Omega|J_X)=|\M(\Omega|J_X)|^2\,\epsilon(\Omega)/I(J_X)$, where
$I(J_X)=\int|\M(\Omega|J_X)|^2\,\epsilon(\Omega){\it d}\Omega$. $\theta_{X}$ is the \xff helicity angle, and 
$\Delta\phi_{X,\pipi}=\phi_{X}-\phi_{\pipi}$ and $\Delta\phi_{X,\jpsi}=\phi_{X}-\phi_{\jpsi}$ are the angles between the $\xff$ decay plane and $\pipi$ or $\jpsi$ decay planes, respectively.
We follow the approach adopted  
in Ref.~\cite{Abulencia:2006ma} 
to predict the matrix elements. 
\begin{figure}[t]
  \begin{center}
    \includegraphics*[height=10cm]{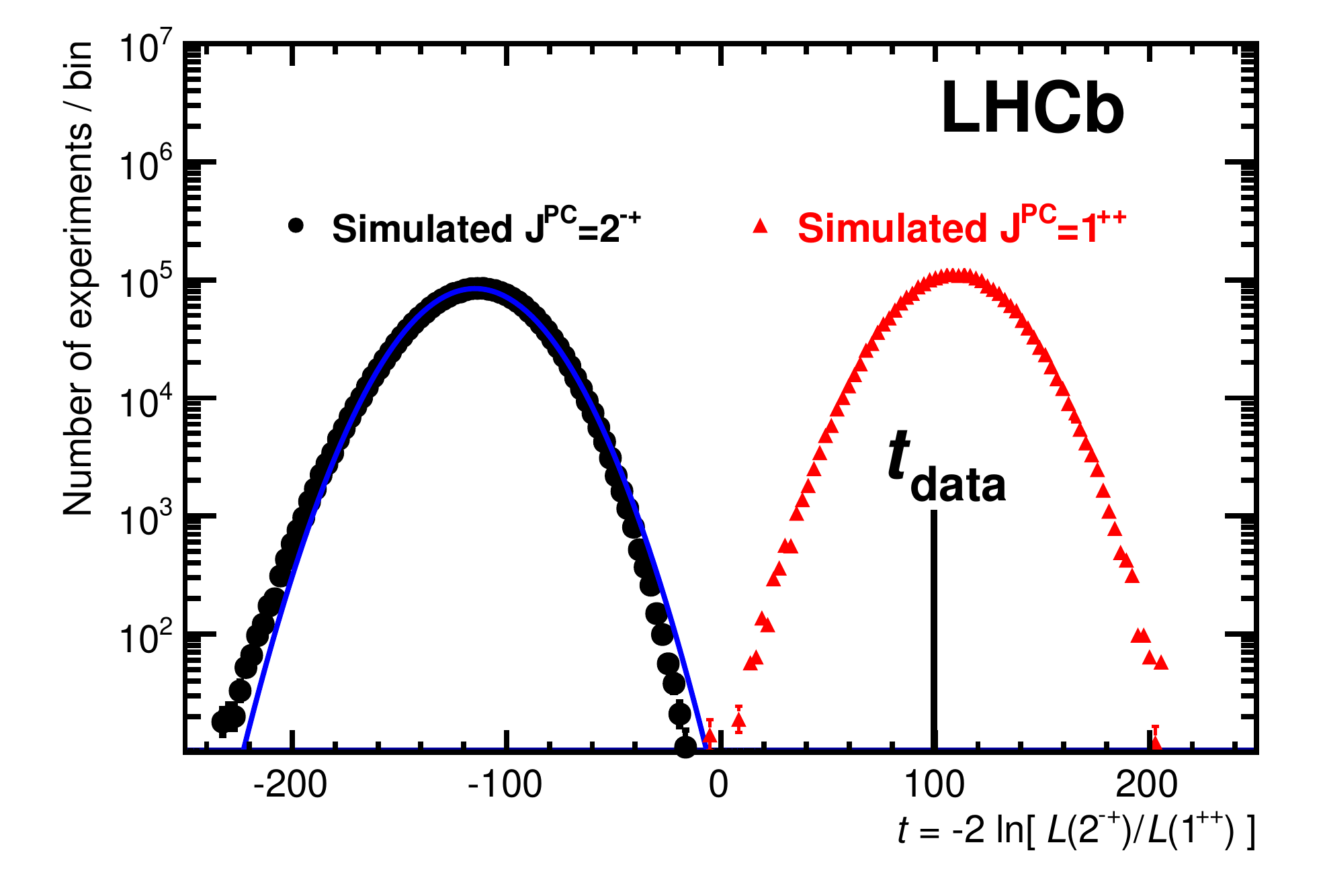} \quad 
  \end{center}
  \vskip-1.1cm\caption{\small
    Distribution of the test statistic $t$ 
    for the simulated experiments 
    with $\jpc=2^{-+}$ 
    (black circles on the left)
    and with $\jpc=1^{++}$ (red triangles on the right).
    A Gaussian fit to the $2^{-+}$ distribution 
    is overlaid (blue solid line).
    The value of the test statistic for the data, $t_{\rm data}$,
    is shown by the solid vertical line. 
  \label{fig:toys}
  }
\end{figure}

We define a test statistic $t=-2\ln[\L(2^{-+})/\L(1^{++})]$. 
The background in the data is subtracted in the log-likelihoods
using the \sPlot\ technique \cite{2005NIMPA.555..356P}.
Positive (negative) values of the test statistic for the data,
$t_{\rm data}$, favor the $1^{++}$ ($2^{-+}$) hypothesis.
The value of the test statistic observed in the data is 
$t_{\rm data}=+99$, favoring the $1^{++}$ hypothesis. 
The value of $t_{\rm data}$ is compared with the distribution 
of $t$ in the simulated experiments
to determine a $p$-value for the $2^{-+}$ hypothesis 
via the fraction of simulated experiments yielding a value of $t>t_{\rm data}$. 
As shown in Fig.~\ref{fig:toys}, 
the distribution of $t$ is
reasonably well approximated by a Gaussian function.
Based on the mean and root mean square 
spread of the $t$ distribution for 
the $2^{-+}$ experiments, this hypothesis is rejected with a significance of $8.4\sigma$. 
Integrating the $1^{++}$ distribution from $-\infty$ to $t_{\rm data}$
gives ${\rm CL}~(1^{++})=34\%$. 
This unambiguously establishes
that $J^{PC}$ for the \xff state
is $1^{++}$. 

~

\section{Study of the $p\overline{p}$ charmonium resonances in \\ {\BppK} decays \cite{BppK}}
The $B^{+} \to p \bar p K^{+}$ decay offers a clean environment to study $c\bar c$ states and
charmonium-like mesons that decay to $p\overline{p}$,
and to search for glueballs or exotic states.
Measurements of intermediate charmonium-like states, such as the \xff, are important to clarify their
nature~\cite{BXK, Brambilla:2010cs} and to determine their partial width to
$p\bar p$, which is crucial
to predict the production rate of these states in dedicated experiments~\cite{Lange:2010dt}. 
\babar and \belle have previously measured the
$\BppK$ branching fraction, including contributions from the $J/\psi$
and $\eta_{c}(1S)$ intermediate states~\cite{Aubert:2005gw, Wei:2007fg}. 
The \lhcb data sample allows the study of
substructures in the $B^{+}\to p\bar p K^{+}$ decays with a sample ten times
larger than those available at previous experiments.

\begin{figure}[t]
\centering
\includegraphics [height=9cm]{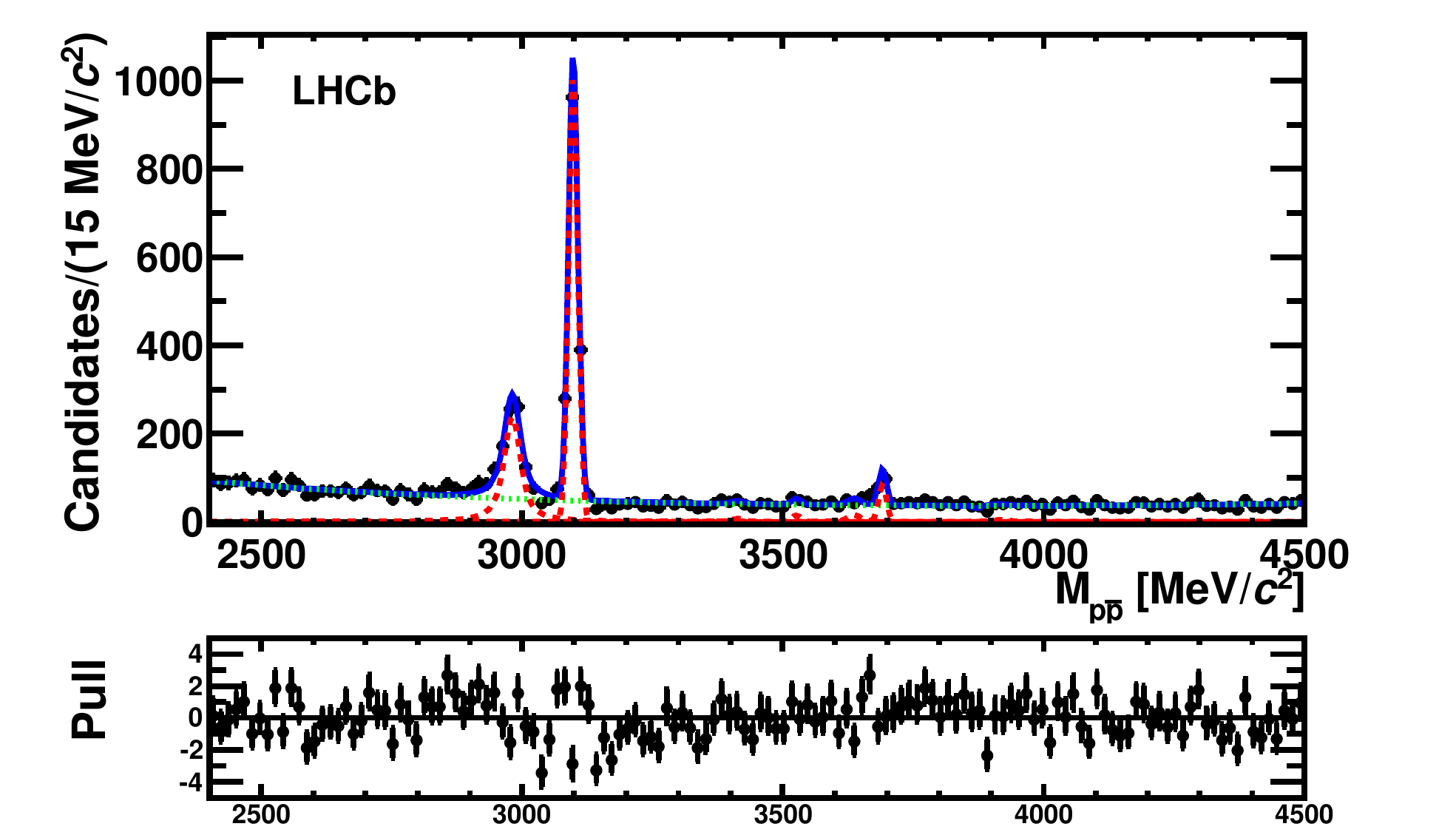}
\caption{
\small Invariant mass distribution of the $p \bar p$ system for 
$B^{+} \to p\bar p K^{+}$ candidates within the $B^{+}$ mass signal window, $\vert M(p\bar p K^{+}) - M_{B^+}\vert < 50 {\rm MeV/}c^{2}$. The dotted lines represent the Gaussian and Voigtian functions (red) and the dashed line the smooth function (green) used to parametrize the signal and the background, respectively.
The bottom plot shows the pulls.
}
\label{fig:cc}
\end{figure}

The signal yields for the charmonium contributions, $B^{+}\to (c\bar c) K^{+} \to p\bar p K^{+}$, are determined by fitting the
$p\bar p$ invariant mass distribution of $B^{+} \to p\bar p K^{+}$ candidates within the $B^{+}$ mass signal window, $\vert M_{p\bar p K^{+}} - M_{B^+}\vert < 50 {\rm MeV/}c^{2}$. 
An unbinned extended maximum likelihood fit to the 
$p\bar p$ invariant mass distribution, shown in Fig.~\ref{fig:cc}, is performed over the mass range $2400-4500~{\rm MeV/}c^{2}$. 
We define the ratio of branching fractions for each resonant ``mode'' as follows:
\begin{equation}
{\cal R}({\rm mode}) =\frac{{\cal B}(B^{+} \to {\rm mode}~K^+\to p\bar p
  K^{+})}{{\cal B}(B^{+}\to J/\psi~K^{+}\to p\bar p K^{+})},
  \end{equation}
where ``mode'' corresponds to the intermediate $\eta_{c}(1S)$, $\psi(2S)$, $\eta_{c}(2S)$,
$\chi_{c0}(1P)$, $h_{c}(1P)$, \xff, or $X(3915)$ states.

Final results for all intermediate modes are given in Ref. \cite{BppK}. The total
branching fraction, its charmless component \mbox{$(M_{p\bar p}
<2.85 {\rm GeV/}c^{2})$} and the branching fractions via the resonant $c\bar c$
states $\eta_{c}(1S)$ and $\psi(2S)$ relative to the decay via a \jpsi
intermediate state are:
\begin{align*}
\frac{{\mathcal B}(\BppK)_{\rm total}}{{\mathcal B}(B^{+} \to J/\psi K^{+} \to p
\bar p K^{+})}=& \, 4.91 \pm 0.19 \, {(\rm
  stat)} \pm 0.14 \, {(\rm syst)},\\
\frac{{\mathcal B}(\BppK)_{\mpp {\rm GeV/}c^{2}}}{{\mathcal B}(B^{+} \to J/\psi K^{+} \to p
\bar p K^{+})}=& \, 2.02 \pm 0.10 \, {(\rm
  stat)}\pm 0.08 \, {(\rm syst)},\\
{\cal R}(\eta_{c}(1S))= & \, 0.578 \pm 0.035 \, {(\rm stat)} \pm 0.027 \, {(\rm syst)},\\
{\cal R}(\psi(2S))=& \, 0.080 \pm 0.012 \, {(\rm stat)} \pm  0.009 \, {(\rm syst)}.
\end{align*}
The branching fractions obtained are compatible with the world average
values~\cite{Beringer:1900zz}. 

We combine our upper limit for $\xff$ with the known value for ${\mathcal B} (B^{+} \to \xff K^{+} ) \times {\mathcal B} (\xff \to
J/\psi \pi^{+} \pi^{-})= (8.6 \pm 0.8) \times 10^{-6}$~\cite{Beringer:1900zz}
to obtain the limit
\begin{equation*}
\frac{{\mathcal B} (\xff \to p \bar p)}{{\mathcal B} (\xff \to J/\psi \pi^{+}
  \pi^{-})}< 2.0\times 10^{-3}.
\end{equation*}
This limit challenges some of the
predictions for the molecular interpretations of the $\xff$ state
and is approaching the
range of predictions for a conventional $\chi_{c1}(2P)$
state~\cite{Chen:2008cg, Braaten:2007sh}.
Using our result and the $\eta_c(2S)$ branching fraction
${\mathcal B} (B^{+} \to \eta_{c}(2S) K^{+})\times
{\mathcal B} (\eta_{c}(2S) \to K \bar K \pi) = (3.4\, ^{+2.3}_{-1.6}) \times 10^{-6}$~\cite{Beringer:1900zz},
a limit of
\begin{equation*} 
\frac{{\mathcal B} (\eta_{c}(2S)
\to p \bar p)}{{\mathcal B} (\eta_{c}(2S) \to K \bar K \pi)} < 3.1 \times 10^{-2}
\end{equation*}
is obtained.

\section{Amplitude analyses of \BnJhh decays \cite{BsJpipi, BsJKK, BJpipi, BJKK}}
Measurement of mixing-induced \CP violation in $\Bsb$ decays is critical for probing physics beyond the Standard Model. 
Final states that are \CP eigenstates with large rates and high detection efficiencies are very useful for such studies. For example, the $\Bsb\rightarrow J/\psi f_0(980)$, $f_0(980)\to\pi^+\pi^-$ decay mode, a \CP-odd eigenstate, was discovered by the LHCb collaboration \cite{Aaij:2011fx} and subsequently confirmed by several experiments \cite{Li:2011pg}. We use only $J/\psi\to\mu^+\mu^-$ decays, so our final state has four charged tracks giving us a high detection efficiency.  LHCb has used this mode to measure the  \CP violating phase $\phi_s$ \cite{LHCb:2011ab}, which complements measurements in the $J/\psi\phi$ final state \cite{LHCb:2011aa,CDF:2011af}. 
It is possible that a larger $\pi^+\pi^-$ mass range could also be used for such studies. In order to fully exploit the $J/\psi \pi^+\pi^-$ final state for measuring \CP violation, it is important to determine its resonant and \CP content. This motivated a ``modified Dalitz plot'' analysis of the \BsJpipi decay.
Modified Dalitz plot analysis differs from a classical Dalitz plot analysis \cite{Dalitz:1953cp} because the $J/\psi$ in our final state has spin-1 and its three decay amplitudes must be considered. We also perform modified Dalitz plot analyses of other \BnJhh decays ($h$ = $\pi$ or $K$). 
\begin{figure}[t]
\vskip -.4cm
\begin{center}
(a) \includegraphics[height=4.7cm]{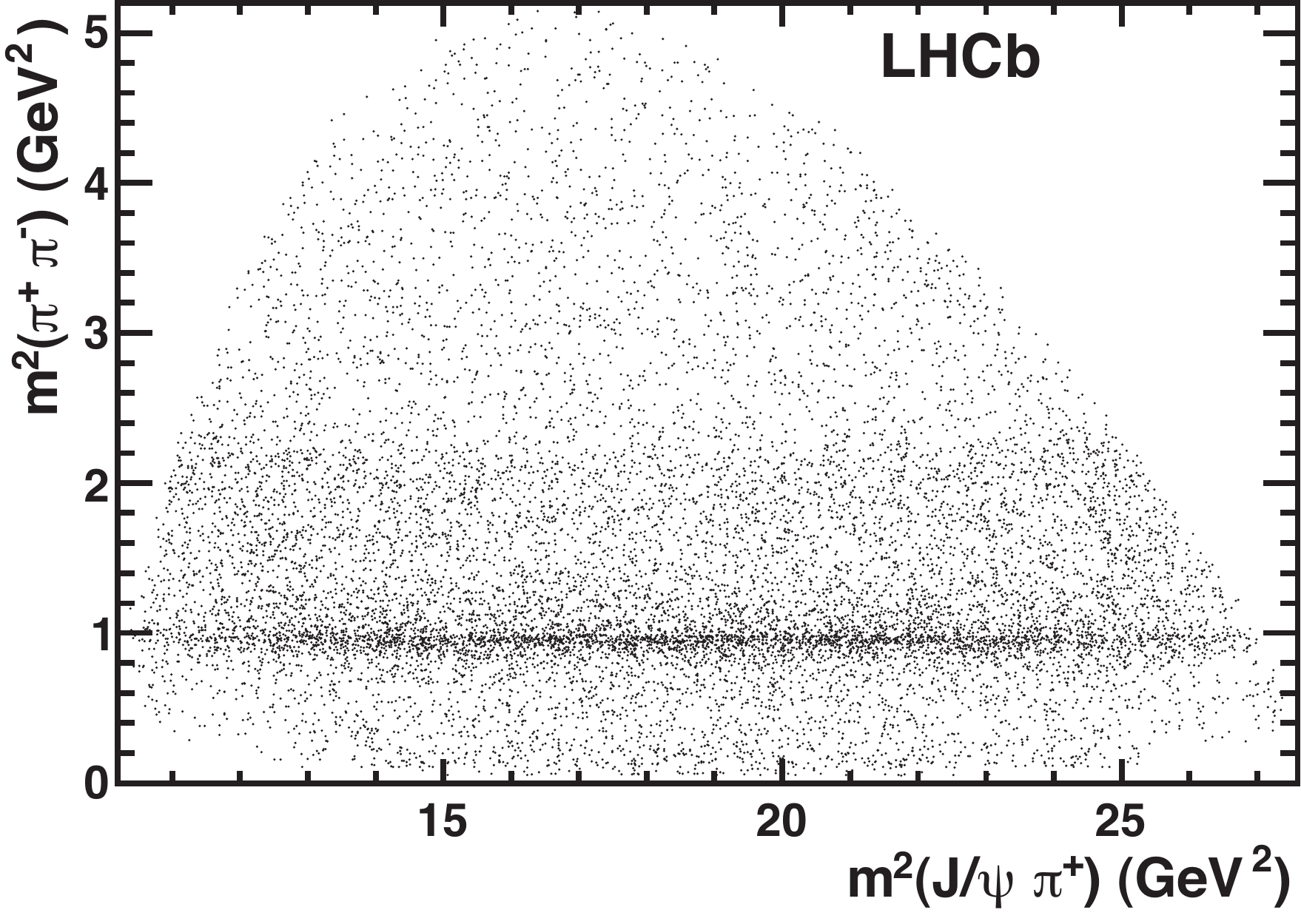}
(b) \includegraphics[height=4.7cm]{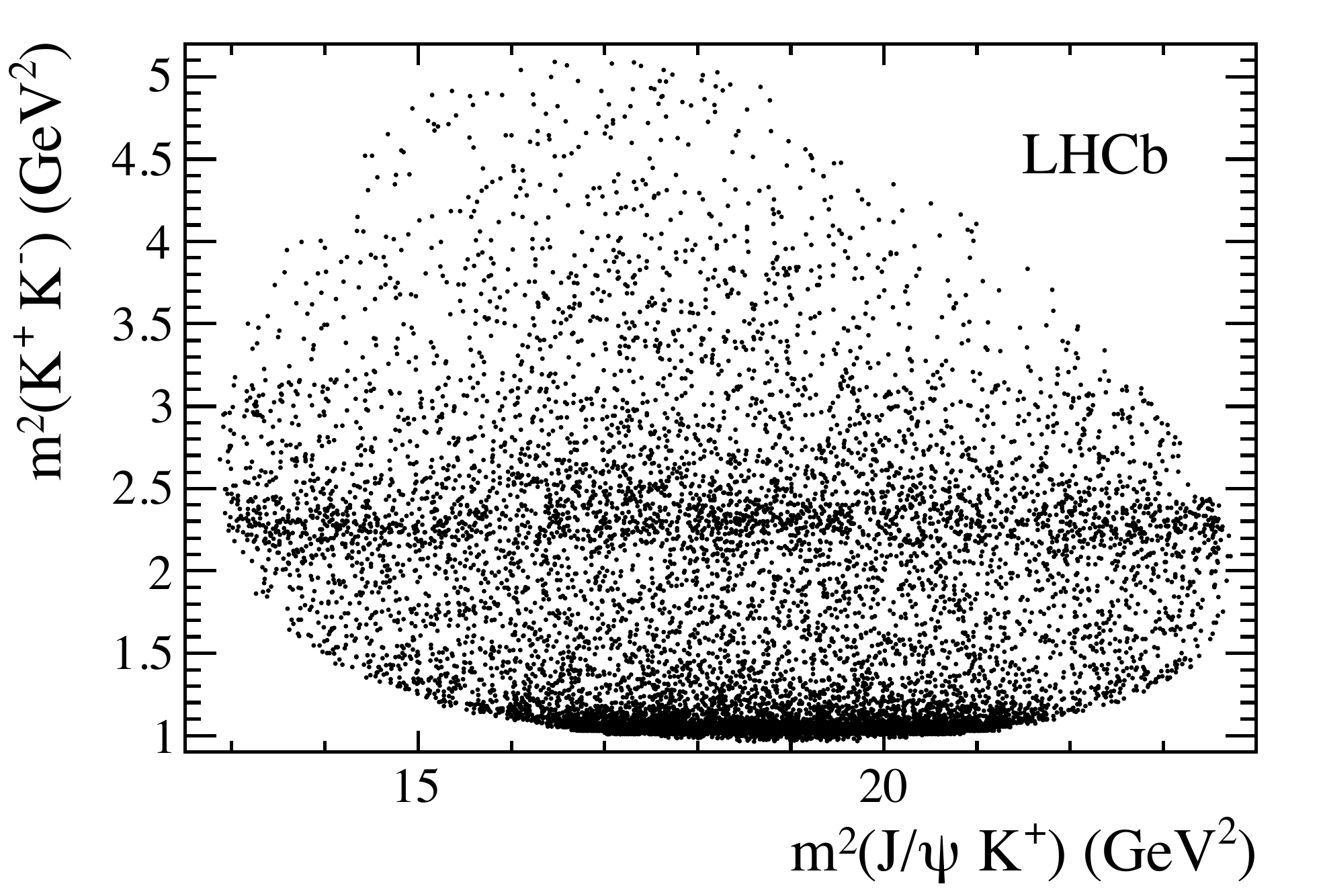}
 \\
(c) \includegraphics[height=4.7cm]{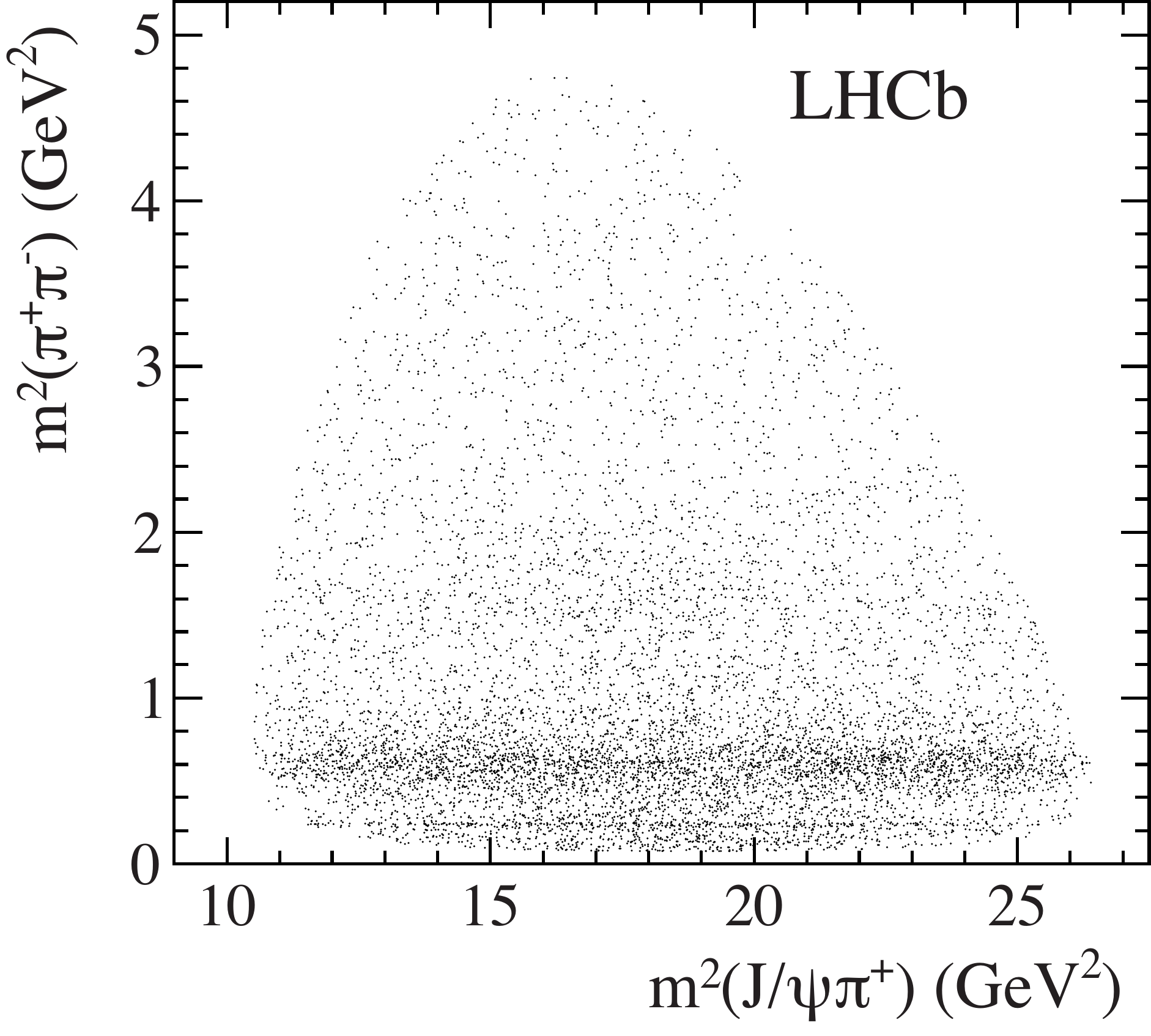}
(d) \includegraphics[height=4.7cm]{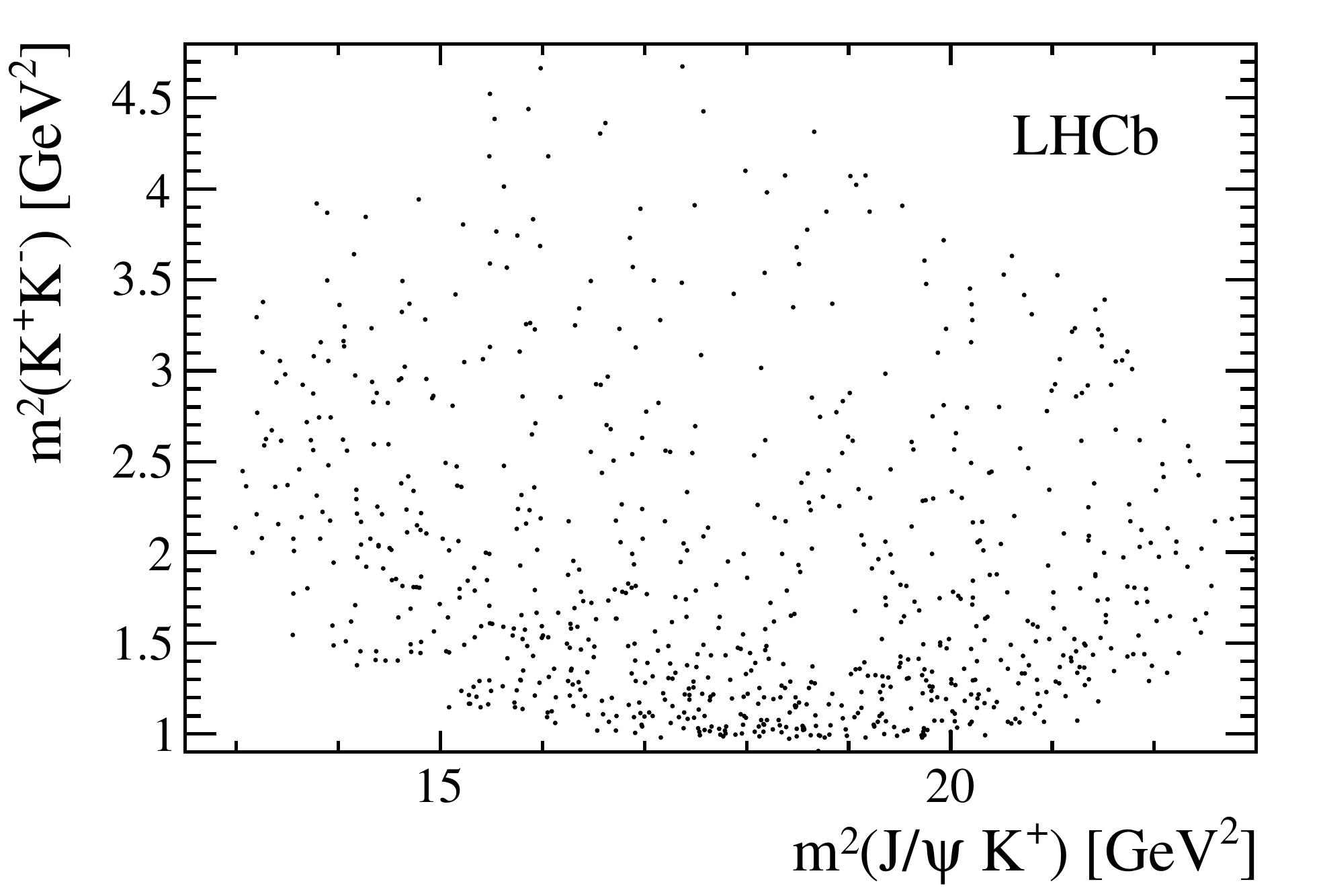}
\end{center}
\vskip -.5cm
\caption{Invariant mass-squared distributions of $\hh$ vs. $J/\psi h^+$ for the decays (a) \BsJpipi, (b) \BsJKK, (c) \BzJpipi, and (d) \BzJKK. Note that in this figure we 
use mass units where we have defined $c=1$.}\label{ddpd}
\end{figure}

In these analyses, we apply a formalism similar to that used in Belle's analysis of $\Bzb\rightarrow K^-\pi^+\chi_{c1}$ decays \cite{Mizuk:2008me}. 
The decay of $\BnJhh$, where $J/\psi\rightarrow \mu^+\mu^-$, can be described by four variables. We choose the invariant mass squared of $J/\psi h^+$ 
($m^2(J/\psi h^+)$), 
the invariant mass squared of $\hh$ 
($m^2(\hh)$), 
the $J/\psi$ helicity angle ($\theta_{J/\psi}$),
and the angle between the $J/\psi$ and $\hh$ decay planes ($\chi$) in the \Bnb rest frame. 
The $\chi$ distribution has little structure, so we analyze the decay process after integrating over $\chi$, which eliminates several interference terms. 
The $m^2(\hh)$ vs. $m^2(J/\psi h^+)$ distributions are shown for the \BnJpipi and \BnJKK decays in Fig.~\ref{ddpd}.
We model the decay with a series of resonant and non-resonant amplitudes. The data are then fitted with the coherent sum of these amplitudes. 

Detailed results of all \BnJhh modified Dalitz plot analyses are available in Refs.~\cite{BsJpipi, BsJKK, BJpipi, BJKK}.
The $\pi^+\pi^-$ system in \BsJpipi is shown to be dominantly in an S-wave state, and the \CP-odd fraction in this \Bsb decay is shown to be greater than 0.977 at 95\% confidence level, meaning that \BsJpipi decays can be used for studies of mixing-induced \CP violation in a large \pipi invariant-mass range.
 In addition, we report the first measurement of the \BsJpipi branching fraction relative to \BsJphi as $(19.79\pm 0.47 \pm 0.52)$\%, where the first uncertainty is statistical and the second is systematic. 
We also report the first observation  of the $\BzJKK$ decay. The branching fraction is determined to be
$\mathcal{B}(\Bdb \to \jpsi \Kp\Km) = (2.53\pm 0.31 \pm 0.19)\times 10^{-6}.$
We also set an upper limit of  $\mathcal{B}(\Bdb \to \jpsi \phi) < 1.9 \times 10^{-7}$ at the 90\% CL, an improvement of about a factor of five with respect to the previous best measurement~\cite{Liu:2008bta}.

\Acknowledgements
We express our gratitude to our colleagues in the CERN
accelerator departments for the excellent performance of the LHC. We
thank the technical and administrative staff at the LHCb
institutes. We acknowledge support from CERN and from the national
agencies: CAPES, CNPq, FAPERJ and FINEP (Brazil); NSFC (China);
CNRS/IN2P3 and Region Auvergne (France); BMBF, DFG, HGF and MPG
(Germany); SFI (Ireland); INFN (Italy); FOM and NWO (The Netherlands);
SCSR (Poland); MEN/IFA (Romania); MinES, Rosatom, RFBR and NRC
``Kurchatov Institute'' (Russia); MinECo, XuntaGal and GENCAT (Spain);
SNSF and SER (Switzerland); NAS Ukraine (Ukraine); STFC (United
Kingdom); NSF (USA). We also acknowledge the support received from the
ERC under FP7. The Tier1 computing centres are supported by IN2P3
(France), KIT and BMBF (Germany), INFN (Italy), NWO and SURF (The
Netherlands), PIC (Spain), GridPP (United Kingdom). We are thankful
for the computing resources put at our disposal by
Yandex LLC (Russia), as well as to the communities behind the multiple open
source software packages that we depend on.

\end{document}